\documentclass[pdflatex,sn-mathphys,Numbered]{sn-jnl}
\usepackage{mathtools}
\usepackage{graphicx}
\usepackage{multirow}
\usepackage{amsmath,amssymb,amsfonts}
\usepackage{amsthm}
\usepackage{mathrsfs}
\usepackage[title]{appendix}
\usepackage{xcolor}
\usepackage{textcomp}
\usepackage{manyfoot}
\usepackage{booktabs}
\usepackage{algorithm}
\usepackage{algorithmicx}
\usepackage{algpseudocode}
\usepackage{listings}
\usepackage{xcolor}
\usepackage{hyperref}

\newcommand{\beq}{\begin{equation}}
\newcommand{\eeq}{\end{equation}}
\newcommand{\beqa}{\begin{eqnarray}}
\newcommand{\eeqa}{\end{eqnarray}}
\newcommand{\be}{\begin{equation}}
\newcommand{\ee}{\end{equation}}
\newcommand{\bea}{\begin{eqnarray}}
\newcommand{\eea}{\end{eqnarray}}

\newcommand{\abs}[1]{\vert#1\vert}
\newcommand{\dd}{{\rm d}}
\newcommand{\e}{{\rm e}}

\newcommand{\ii}{{\rm i}}
\renewcommand{\max}{{\rm max}}
\newcommand{\mean}[1]{\langle#1\rangle}
\newcommand{\prob}{\mathbb{P}}
\renewcommand{\th}{{\theta}}
\newcommand{\Int}{\mathop{{\rm Int}}}

\renewcommand{\Re}{\mathop{{\rm Re}}}

\begin{document}

\title{New insights into the distribution of the topmost gap in random walks and L\'evy flights}

\author*[]{\fnm{Claude} \sur{Godr\`eche*}}\email{claude.godreche@ipht.fr}

\author[]{\fnm{Jean-Marc} \sur{Luck}}\email{jean-marc.luck@ipht.fr}

\affil[]
{\orgdiv{Universit\'e Paris-Saclay, CEA, CNRS},
\orgname{Institut de Physique Th\'eorique},
\postcode{91191}
\city{Gif-sur-Yvette},
\country{France}}

\abstract{
Building upon the knowledge of the distribution of the first positive position
reached by a random walker starting from the origin,
one can derive new results on the statistics of the gap between
the largest and second-largest positions of the walk,
and recover known ones in a more direct manner.}


\maketitle

\tableofcontents

\section{Introduction}

Extreme-value theory and order statistics have been the subject of extensive study in probability theory~\cite{lamperti,DN03,Resnick87,Leadbetter83,deHaan06}, 
statistical physics~\cite{BouchaudMezard97,MOR11,castellana,fortin,msbook}, astrophysics~\cite{TR77}, environmental sciences~\cite{KPN02}, 
finance~\cite{EKM97,BouchaudPotters03}, reliability theory~\cite{barlow}, and other fields~\cite{gumbel,galambos}.
Classical results mostly concern independent and identically distributed random variables~\cite{ABN92, DN03, Reiss89}.

The study of order statistics for one-dimensional random walks was initiated by Pollaczek~\cite{Pol52}, Wendel~\cite{Wendel60}, and others.
More broadly, the fluctuation theory of random walks, developed in the 1950s and 1960s by Sparre Andersen, Spitzer, Baxter, Feller, and Wendel, describes the distributions of extrema and the random times at which they are first or last attained.

Recent work~\cite{SM12} on random walk order statistics has focused on the spacings between the successive ordered positions of the walker.
For symmetric walks with continuous increments, the authors derive new limit laws and asymptotic expressions.
In particular, when the step distribution is symmetric with finite variance, the expected spacings converge at long times to explicit integral formulas involving the Fourier transform of the step density.
This question is further explored in~\cite{pitman}, which also examines the long-time behavior of these spacings%
\footnote{We refer to the latter reference for more details on the historical aspects of random walk extreme order statistics.
The beginning of the present introduction borrows from it.}.

Here, we establish new results concerning the topmost gap $Z$, defined as the distance 
between the largest and second-largest positions of a formally infinitely long walk.
Our approach bypasses the more intricate computations required in earlier
studies~\cite{mounaix1,mounaix2},
which aimed to derive the joint distribution of $Z$ and the time interval
between the largest and second-largest positions.

This paper builds upon~\cite{fpp}, in which we carried out a comprehensive study of the
distribution of the first positive position reached by a random walk starting from the origin,
\be
x_n=\eta_1+\cdots+\eta_n, \qquad x_0=0,
\ee
where the steps $\eta_1,\eta_2,\dots$ are independent and identically distributed random variables
with a common probability density function $\rho(x)$.
The latter is assumed to be symmetric and continuous,
allowing the random walk to exhibit either diffusive behaviour (finite variance, $\mean{\eta^2}=2D$)
or L\'evy flight dynamics (infinite variance).
Let $N\ge1$ denote the first time at which the walker's position becomes positive,
and let $H=x_N$ be this first positive position (also called the first ladder height).
The probability density function of the random variable $H$ is defined by
\beq\label{eq:fH}
f_H(x)=\frac{\dd}{\dd x}\prob(H<x).
\eeq
In~\cite{fpp}, we derived quantitative estimates for the asymptotic tail behaviour of $f_H(x)$, which led us to classify step distributions into three distinct categories:
superexponential, exponential, and subexponential.
We also performed a detailed analysis of the case where the step distribution follows a stable law and computed the moments of $H$ for a Gaussian walk.
Moreover, we obtained explicit expressions for $f_H(x)$ in specific cases,
notably for the family of symmetric Erlang distributions.
The notations of~\cite{fpp} will be used throughout.

Our main results concerning the topmost gap $Z$ between the largest and second-largest positions of a very long random walk
are as follows:
\begin{enumerate}
\item An interpretation of $Z$
as the smaller of two independent copies of $H$, the first positive position of the walk,
leading to a direct derivation of an expression for the gap density $f_Z(x)$ in terms of $f_H(x)$.
\item A systematic characterisation of the asymptotic tail behaviour of $f_Z(x)$,
for the three different classes of step distributions mentioned above.
\item An approach to the computation of the moments of $Z$.
\item A detailed study of the family of symmetric Erlang distributions,
including the explicit determination of $f_Z(x)$ and the study of its moments and of other characteristics.
\end{enumerate}

\section{Distribution of the gap}

\begin{figure}[!htbp]
\begin{center}
\includegraphics[angle=0,width=.6\linewidth,clip=true]{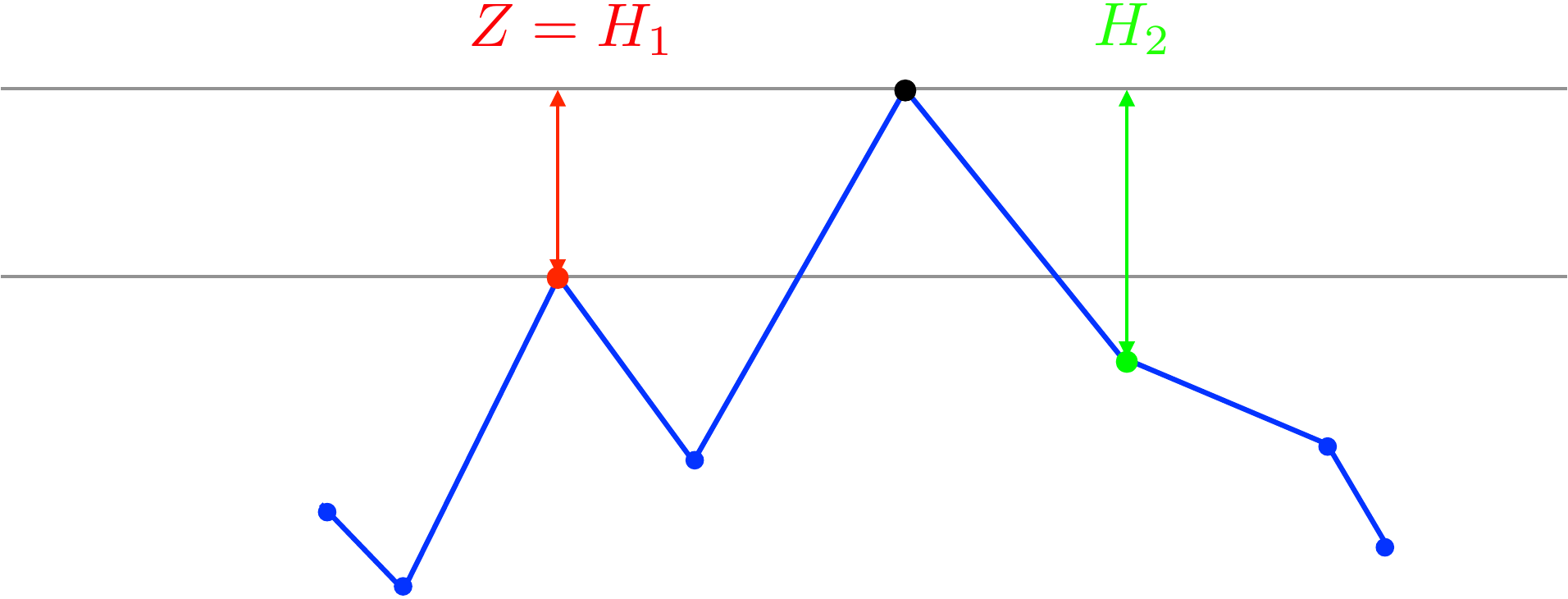}
\includegraphics[angle=0,width=.6\linewidth,clip=true]{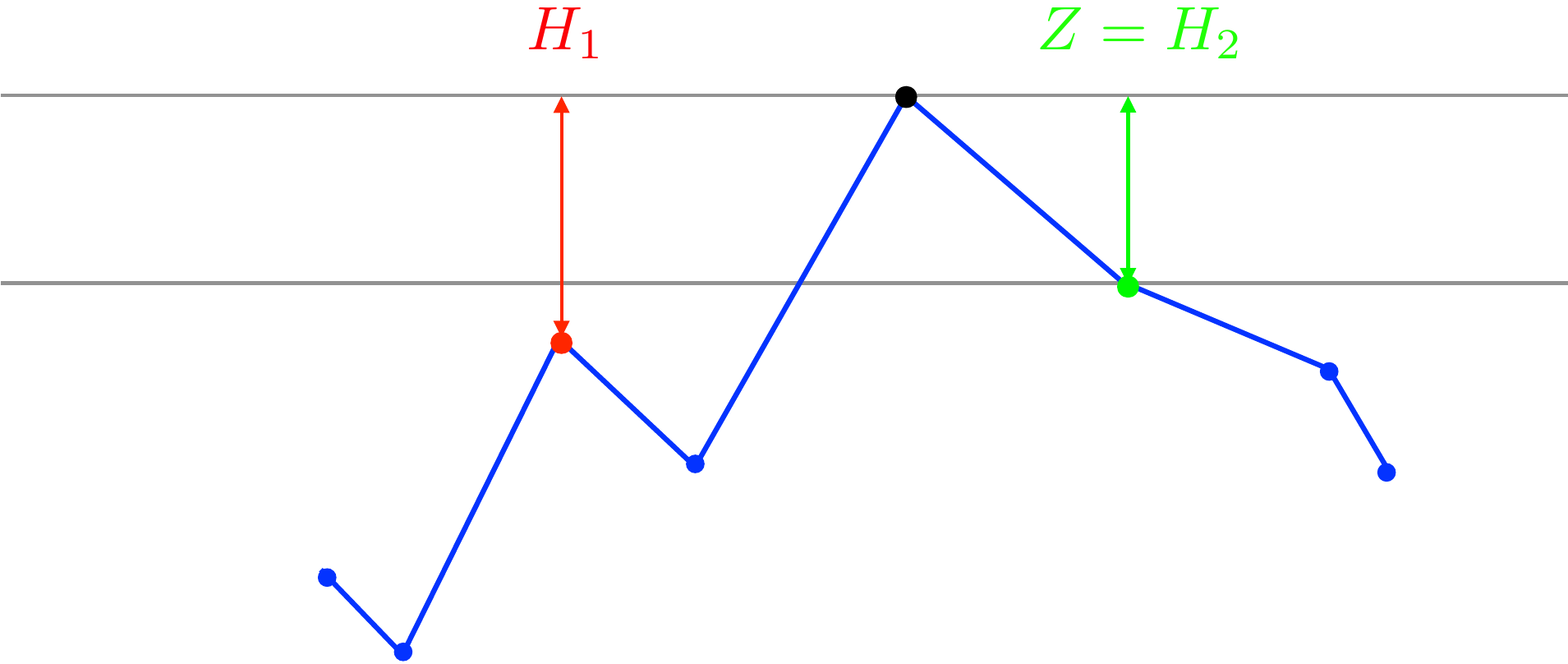}
\caption{Local structure of a formally infinite walk in the vicinity of its absolute maximum.
The gap $Z$ is the smaller of two independent copies of $H$, the first positive position of the walk,
obtained when reading the walk either from left to right or from right to left (see text).
}
\label{fig:gap}
\end{center}
\end{figure}

The key observation of our approach is that $Z$ can be viewed as the smaller of two independent copies of $H$, the first positive position of the walk:
\be
Z = \min(H_1, H_2).
\label{zmin}
\ee
This interpretation is illustrated in figure~\ref{fig:gap}.
In each panel, the black point represents the absolute maximum,
i.e., the highest point of a formally infinite walk.
The distance $H_1$ corresponds to the first positive position of the walk when
read from left to right, starting from the red point.
In this case, this position is reached after two steps.
Similarly, $H_2$ represents the first positive position when the walk is
read from right to left, starting from the green point.
Here, this position is reached after a single step.
In the first panel, where $H_1<H_2$, the red point is the second-highest, so $Z=H_1$.
In the second panel, where $H_2<H_1$, the green point is the second-highest, giving $Z=H_2$.%
\footnote{Notice that the red point is a local maximum of the walk, whereas the green one is not.
In general, the second-largest position is a local maximum
when it is separated from the largest position by at least two steps.
This event occurs with probability $1/2$~\cite{fpp}.}
To the best of our knowledge, the relation~\eqref{zmin} was not previously known in the fluctuation theory of random walks.

It follows from~(\ref{zmin}) that
\beq
\prob(Z>x)=\prob(H>x)^2,
\label{evs}
\eeq
entailing the following relation between the densities of these two random variables
\beq\label{eq:fzh}
f_Z(x)=2 f_H(x) \prob(H>x)
= 2 f_H(x) \int_x^\infty\dd y\,f_H(y).
\eeq
The density of $f_H(x)$ is given in Laplace space by
\beq\label{eq:identity}
\hat f_H(p)=1-\frac{1}{\hat g(p)},
\eeq
where
\beq
\label{eq:Poll-S}
\hat g(p)
=\int_0^\infty\dd x\, \e^{-p x} g(x)
=\exp\left(-\frac{p}{\pi}\int_0^\infty\frac{\dd q }{p^2+q^2}\ln(1-\tilde \rho(q))\right),
\eeq
and
\be
\tilde \rho(q)=\int_{-\infty}^{\infty}\dd x\, \e^{\ii q x} \rho(x)
\ee
is the Fourier transform of the step density $\rho(x)$.
The expression~\eqref{eq:Poll-S} is known as the Pollaczek-Spitzer
formula~\cite{spitzer,spitzer1,spitzer2,spitzer3,pollaczek}.
In this expression,
\be
g(x)=\delta(x)+\sum_{n\ge1}g_n(x),
\ee
where
\be
g_n(x) \dd x=\prob(x_1>0,\dots,\ x_{n-1}>0,\ x<x_n<x+\dd x).
\ee
We also have
\be
g(x)=\frac{\dd}{\dd x} G(x),
\ee
where
$G(x)$ is
the expected number of records in the interval $(0,x)$ (counting the record at the origin), or
else, the expected number of visits to the same interval such that $x_k>0$ for all $k=1,\dots,n$.
It is also the solution of the homogeneous Wiener-Hopf equation with kernel $\rho(x)$,
\beq\label{WH}
G(x)=\int_{0}^{\infty}\dd y\, G(y)\rho(x-y),
\eeq
with boundary condition $G(0)=1$~\cite{fpp}.

As an immediate consequence of~(\ref{eq:fzh}), the value of $f_Z(x)$ at the origin is given by $f_Z(0)=2\omega$,
where~\cite{fpp}
\beq\label{eq:omegaint}
\omega=f_H(0)=-\frac{1}{\pi}\int_0^\infty\dd q\,\ln(1-\tilde\rho(q)).
\eeq

For diffusive walks,~\eqref{eq:fzh} can be rewritten as
\beq
f_Z(x)=2 \sqrt{D} f_H(x) f_E(x),
\label{zhe}
\eeq
where the excess length $E$,
with distribution given by
\be
f_E(x)=\frac{1}{\sqrt{D}}\int_x^\infty\dd y\,f_H(y),
\ee
is the stationary limit of the overshoot $E_x$ of the random walk over the `barrier' located at
height $x$~\cite{fpp}.
In Laplace space, this reads
\beq\label{eq:hatfE}
\hat f_E(p)=\frac{1-\hat f_H(p)}{p\sqrt{D}}=\frac{1}{p\sqrt{D}\,\hat g(p)},
\eeq
where the last equality is due to~\eqref{eq:identity}.

Equations~\eqref{eq:identity} and~\eqref{eq:Poll-S} form the basis of the method for determining
$f_H(x)$---hence of $f_Z(x)$ using~\eqref{eq:fzh} (or~(\ref{zhe})).
Alternatively, one could work in direct space using the following two formulas:
\beq\label{eq:fHx}
f_H(x)=\int_0^{\infty}\dd y\,g(y) \rho(x+y),
\eeq
\beq\label{eq:Grelation}
\prob(H>x)=\int_0^{\infty}\dd y\,G(y) \rho(x+y).
\eeq
Equation~\eqref{eq:fHx} was noted in~\cite{gmsprl} (see also~\cite{fpp,revue}), while
\eqref{eq:Grelation} represents a novel result.
However, analysing $f_H(x)$---and therefore $f_Z(x)$---is more conveniently performed
using~\eqref{eq:identity} and~\eqref{eq:Poll-S}, as demonstrated in~\cite{fpp}.

In~\cite{mounaix1,mounaix2},
it was found that $f_Z(x)=2 I_1(x) I_2(x)$,
where $I_1(x)$ and $I_2(x)$ correspond to the integrals on the right-hand sides of~\eqref{eq:fHx}
and~\eqref{eq:Grelation}, respectively.
However, their interpretations in terms of $f_H(x)$ and $\prob(H>x)$ were not provided, and the
derivation of this result was significantly more intricate.

\section{Asymptotic tail behaviour}

The asymptotic behaviour of the tail of $f_Z(x)$ can be directly deduced from that of $f_H(x)$
using~(\ref{evs}) or~(\ref{eq:fzh}), based on the study carried out in~\cite{fpp} with the help
of the Wiener-Hopf factorisation identity.
In Fourier space, this identity is given by~\cite{feller2}
\beq\label{eq:WHF}
(1-\tilde f_H(q))(1-\tilde f_H(-q))=1 -\tilde \rho(q).
\eeq
It extends to the Laplace domain as
\beq\label{eq:WHL}
(1-\hat f_H(p))(1-\hat f_H(-p))=1-\hat \rho(p),
\eeq
provided that $\rho(x)$ decays at least exponentially.
Along the lines of~\cite{fpp},
three classes of step distributions must be treated separately.

References~\cite{mounaix1,mounaix2} address the same question,
namely the asymptotic tail behaviour of $f_Z(x)$,
but rely on the analysis of the expression $f_Z(x)=2I_1(x)I_2(x)$.
The approach presented here is considerably simpler.

\subsection{First class}

The first class consists of superexponentially decaying step distributions.
Examples include the uniform distribution (finite support) and the Gaussian distribution (infinite
support).
The Laplace transform $\hat \rho(p)$ is analytic in the entire complex $p$-plane.
The Wiener-Hopf factorisation identity~\eqref{eq:WHL} yields~\cite{fpp}
\be
f_H(x)\approx\rho(x),
\ee
where $\approx$ denotes asymptotic equivalence,
as $x$ approaches the upper edge of the common support of $\eta$ and $H$,
whether finite or infinite.
Using~(\ref{eq:fzh}), we thus obtain
\be
f_Z(x)\approx 2\rho(x)\prob(\eta>x)=2\rho(x)\int_x^\infty\dd y\,\rho(y).
\ee
For instance, for the Gaussian distribution $\rho(x)={\e^{-x^2/4}}/{\sqrt{4\pi}}$, this reads
\be
f_Z(x)\approx \frac{\e^{-x^2/2}}{\pi x}.
\ee

\subsection{Second class}

The second class consists of exponentially decaying step distributions:
\be
\rho(x)\sim\e^{-b\abs{x}}.
\ee
In this case, the Laplace transform $\hat \rho(p)$ is analytic in the strip $\abs{\Re p\,}<b$.
The Wiener-Hopf factorisation identity~\eqref{eq:WHL} yields~\cite{fpp}
\be
f_H(x)\approx K \rho(x),
\qquad K=\frac{1}{1-\hat f(b)}=\hat g(b),
\label{eq:K}
\ee
entailing the result
\be
f_Z(x)\approx 2K^2\rho(x)\prob(\eta>x).
\ee
The constant $K$ depends on parameters of the step distribution (see, e.g.,~(\ref{qties})).

\subsection{Third class}

The third class consists of subexponentially decaying step distributions, that is, those that decay
more slowly than any exponential function.
Henceforth, we focus on those with a power-law decay,
\be
\rho(x)\approx\frac{c}{x^{1+\th}}.
\ee

\begin{enumerate}

\item[1.]
For diffusive walks ($\theta>2$), such that $D$ is finite, we have~\cite{fpp}
\beq\label{eq:fHdiff}
f_H(x)\approx\frac{a}{x^\theta},\qquad
a=\frac{c}{\theta\sqrt{D}},
\eeq
and thus
\be
f_Z(x)\approx\frac{A}{x^{2\theta-1}},\qquad
A=\frac{2a^2}{\theta-1}=\frac{2c^2}{\theta^2(\theta-1)D}.
\ee
The gap distribution exhibits a steeper power-law decay than the step distribution,
since $2\theta-1>1+\theta$ for $\theta>2$.
\item[2.]
For L\'evy flights ($0<\theta<2$), we have~\cite{fpp}
\beq\label{eq:fHLev}
f_H(x)\approx\frac{a}{x^{1+\theta/2}},\qquad
a=R(\theta)\sqrt{c},
\eeq
with
\be
R(\theta)=\Gamma(1+\theta/2)\left(\frac{\sin(\pi\theta/2)}{\pi\Gamma(1+\theta)}\right)^{1/2},
\ee
and thus
\be
f_Z(x)\approx\frac{A}{x^{1+\theta}},\qquad
A=\frac{4a^2}{\theta}=S(\theta)c,
\ee
with
\be
S(\theta)=\frac{4R(\theta)^2}{\theta}=\frac{\Gamma(\theta/2)}{\Gamma(\theta)\Gamma(1-\theta/2)}.
\ee
The step and gap distributions now have the same decay exponent,
so that they are asymptotically proportional to each other:
\beq
f_Z(x)\approx S(\theta) \rho(x).
\eeq
The universal proportionality factor $S(\theta)$ decreases from $S(0)=2$ through $S(1)=1$
to $S(\theta)\approx(2-\theta)/2$ as $\theta\to2$.

\item[3.]
In the marginal case where $\theta=2$,
i.e., $\rho(x)\approx c/\abs{x}^3$,
the distribution of $H$ exhibits a logarithmic correction of the form~\cite{fpp}
\be
f_H(x)\approx\frac{1}{2x^2}\left(\frac{c}{\ln x}\right)^{1/2}.
\ee
We have therefore
\be
f_Z(x)\approx\frac{c}{2x^3\ln x}.
\ee
\end{enumerate}

The preceding results on the tail behaviour of $f_H(x)$ and $f_Z(x)$ can be summarised as follows:
\be
f_H(x)\approx\frac{a}{x^{1+\sigma_H}},\qquad
f_Z(x)\approx\frac{A}{x^{1+\sigma_Z}},
\ee
with
\be
\sigma_Z=2\sigma_H=\left\{
\begin{array}{cl}
\theta\quad & (\theta<2),\\
2(\theta-1)\quad & (\theta>2),
\end{array}
\right.
\ee
and
\beq
A=\frac{2a^2}{\sigma_H}.
\eeq
These expressions are consistent with~(\ref{evs}).
They have the following consequences for the gap moments.
The mean gap $\mean{Z}$ is finite for $\theta>1$,
whereas $\mean{Z^2}$ is finite for $\theta>2$,
$\mean{Z^3}$ for $\theta>5/2$,
$\mean{Z^4}$ for $\theta>3$,
and so on.
In general, the moments $\mean{Z^k}$ are finite for $k=1,\dots,k_\max$, with $k_\max=\Int(\sigma_Z)$.

\section{Moments}

We first recall a few results concerning the moments of the first positive position~$H$~\cite{fpp}.
We have
\beq
\mean{H}=\sqrt{D},\qquad
\mean{H^2}=2\sqrt{D}\,\ell.
\label{h12res}
\eeq
The first moment is finite for all diffusive walks ($\mean{\eta^2}=2D$ finite).
The second moment involves the extrapolation length $\ell$, which reads
\beq
\ell=-\frac{1}{\pi}\int_0^\infty\frac{\dd q}{q^2}\ln\frac{1-\tilde\rho(q)}{D q^2},
\label{eq:ell}
\eeq
and converges whenever $\mean{\abs{\eta}^3}$ is finite.
Higher-order moments $\mean{H^k}$ can be expressed in terms of the cumulants $c_k$ of the
stationary excess length $E$.
Even cumulants $c_{2m}$ have simple expressions, whereas odd cumulants $c_{2m+1}$ are given by
integral representations generalising the expression~(\ref{eq:ell}) for $\ell=c_1$~\cite{fpp}.

The relation~(\ref{zmin}), however, suggests that there is no simple relationship between the
moments of $Z$ and $H$ in general, making the evaluation of the gap moments a more challenging
task, which we shall now address.
For now, we assume that the step distribution decays at least exponentially, ensuring that the
Laplace transform $\hat\rho(p)$ is analytic at least in a strip $\abs{\Re p\,}<b$.

\subsection{First moment}

The first moment of the gap reads (see~(\ref{eq:fzh}))
\beq
\mean{Z}=2\int_0^\infty\dd x\,x\,f_H(x) \int_x^\infty\dd y\,f_H(y).
\eeq
Replacing $f_H(x)$ by its Laplace integral representation
\beq
f_H(x)=\int\frac{\dd p}{2\pi\ii}\,\e^{-px}\,\hat f_H(-p)\qquad(0<\Re p<b),
\eeq
and similarly for $f_H(y)$, we have
\beq
\mean{Z}=\int\frac{\dd p}{2\pi\ii}\,\hat f_H(-p)\int\frac{\dd r}{2\pi\ii}\,\hat f_H(-r)
\left(2\int_0^\infty\dd x\,x\,\e^{-px}\int_x^\infty\dd y\,\e^{-ry}\right).
\eeq
The parenthesis equals $2/(r(p+r)^2)$,
which simplifies to $1/(pr(p+r))$ after symmetrisation with respect to $p$ and $r$, yielding
\beq
\mean{Z}=\int\frac{\dd p}{2\pi\ii}\,\hat f_H(-p)\int\frac{\dd r}{2\pi\ii}\,\hat
f_H(-r)\,\frac{1}{pr(p+r)}.
\label{z1sym}
\eeq
Integrating over $r$ amounts to adding up the residues of the poles at $r=0$ and $r=-p$.
We thus obtain
\beq
\mean{Z}=\int\frac{\dd p}{2\pi\ii\,p^2}\,\hat f_H(-p)(1-\hat f_H(p)).
\eeq
Adding the vanishing integral
\beq
\int\frac{\dd p}{2\pi\ii\,p^2}(1-\hat f_H(p))=0,
\eeq
and applying the Wiener-Hopf factorisation identity~(\ref{eq:WHL}), we obtain
\beq
\mean{Z}=-\int\frac{\dd p}{2\pi\ii\,p^2}(1-\hat f_H(-p))(1-\hat f_H(p))
=-\int\frac{\dd p}{2\pi\ii\,p^2}(1-\hat\rho(p)).
\label{z1int}
\eeq
Evaluating this integral gives 
\beq
\mean{Z}=\frac{\mean{\abs{\eta}}}{2}.
\label{z1res}
\eeq
This is consistent with the remark made in~\cite{pitman}, that, for any symmetric distribution of $\eta$, the following identity holds:
\beq\label{eq:pitman}
\mean{\abs{\eta}}= 2 \int_{0}^\infty\frac{\dd q}{\pi q^2} (1-\tilde\rho(q)).
\eeq
Thus, $\mean{Z}$ equals the integral in~\eqref{eq:pitman}, which is the counterpart in Fourier space of the Laplace integral in the right-hand side of~\eqref{z1int}, 
recovering a result from~\cite{mounaix1,mounaix2}.
The expressions~(\ref{h12res}) for $\mean{H}$ and~(\ref{z1res}) for $\mean{Z}$ share a similar simplicity, both
involving moments of the step length $\abs{\eta}$.
The inequality $\mean{\abs{\eta}}^2\le\mean{\eta^2}$ translates to
\beq
\mean{Z}\le\frac{\mean{H}}{\sqrt2}.
\label{zhineq}
\eeq

\subsection{Second moment}

The analogue of~(\ref{z1sym}) for the second moment of the gap is given by
\beq
\mean{Z^2}=\int\frac{\dd p}{2\pi\ii}\,\hat f_H(-p)\int\frac{\dd r}{2\pi\ii}\,\hat
f_H(-r)\,\frac{2}{pr(p+r)^2}.
\eeq
Integrating over $r$ still amounts to summing the residues of the poles at $r=0$ and $r=-p$.
The latter is now a double pole,
whose residue involves a derivative with respect to $p$,
which we denote by an accent.
This gives
\beq
\mean{Z^2}=2\int\frac{\dd p}{2\pi\ii\,p^3}\,\hat f_H(-p)(1-\hat f_H(p)+p\hat f'_H(p)).
\label{z2res1}
\eeq
Using~(\ref{eq:hatfE}), this expression can be rewritten as
\beq
\mean{Z^2}=-2\sqrt{D}\int\frac{\dd p}{2\pi\ii\,p}\,\hat f_H(-p)\hat f'_E(p).
\label{z2res2}
\eeq
The analyticity assumption made above can be relaxed as follows.
For step distributions with $\theta>3$,
the extrapolation length $\ell$ defined in~(\ref{eq:ell}) is finite,
and so $\hat f'_E(0)=-\mean{E}=-\ell$.
As a result,~(\ref{z2res2}) can be rewritten in terms of Fourier transforms
by carefully taking the limit $p\to\ii q$.
We thus obtain
\beq
\mean{Z^2}=\ell\sqrt{D}
+\frac{\sqrt{D}}{\pi}\int_{-\infty}^\infty\frac{\dd q}{q}\,(\tilde f_H(-q)\tilde f'_E(q)-\ii\ell).
\label{z2res3}
\eeq
For step distributions such that $2<\theta<3$, $\mean{Z^2}$ is finite, although $\ell$ is divergent.
Obtaining an integral expression for $\mean{Z^2}$ in that case requires greater care.
We shall not elaborate on this.

\subsection{Higher moments and Laplace transform}

Higher moments of the gap distribution can be obtained following the same approach as above.
It is convenient to wrap them up into the Laplace transform
\beq
\hat f_Z(s)=\mean{\e^{-sZ}}=\int_0^\infty\dd x\,\e^{-sx}\,f_Z(x).
\eeq
The analogue of~(\ref{z1sym}) is
\beq
\hat f_Z(s)=\int\frac{\dd p}{2\pi\ii}\,\hat f_H(-p)\int\frac{\dd r}{2\pi\ii}\,\hat
f_H(-r)\,\frac{2}{r(p+r+s)}.
\eeq
Integrating over $r$ amounts to adding up the residues of the poles at $r=0$ and $r=-(p+s)$.
We thus obtain
\beq
\hat f_Z(s)=\int\frac{\dd p}{2\pi\ii}\,\hat f_H(-p)(1-\hat f_H(p+s))\,\frac{2}{p+s}.
\label{fzres1}
\eeq
Using again~(\ref{eq:hatfE}), this expression can be rewritten as
\beq
\hat f_Z(s)=2\sqrt{D}\int\frac{\dd p}{2\pi\ii}\,\hat f_H(-p)\hat f_E(p+s).
\label{fzres2}
\eeq
Finally, the analyticity hypothesis can by released by taking the limit $p\to\ii q$, yielding
\beq
\hat f_Z(s)=\frac{\sqrt{D}}{\pi}\int_{-\infty}^\infty\dd q\,\tilde f_H(-q)\hat f_E(s+\ii q).
\label{fzres3}
\eeq

In general, extracting more explicit results from the quadratic integral
expressions~(\ref{z2res1}),~(\ref{z2res2}),~(\ref{z2res3}) for $\mean{Z^2}$
and~(\ref{fzres1}),~(\ref{fzres2}),~(\ref{fzres3}) for $\hat f_Z(s)$ is virtually impossible.
This observation confirms our expectation that evaluating higher moments of $Z$ is indeed a
difficult task.

\section{Symmetric Erlang distributions}

For this family of distributions, the step density is given by
\beq
\rho(x)=\frac{\abs{x}^{M-1}\e^{-\abs{x}}}{2(M-1)!},
\label{rhon}
\eeq
where the Erlang parameter $M\ge1$ is integer.
In Laplace space,~\eqref{rhon} reads
\beq
\hat\rho(p)=\frac12\left(\frac{1}{(1+p)^M}+\frac{1}{(1-p)^M}\right).
\label{laprhon}
\eeq
We have $\mean{\abs{\eta}}=M$ and $\mean{\eta^2}=2D=M(M+1)$, and therefore
\beq
\mean{H}=\sqrt{\frac{M(M+1)}{2}},\qquad
\mean{Z}=\frac{M}{2},\qquad
\frac{\mean{Z}}{\mean{H}}=\sqrt{\frac{M}{2(M+1)}}.
\label{momsn}
\eeq
The latter ratios converge to the bound~(\ref{zhineq}) in the limit of large $M$.

\begin{figure}[!htbp]
\begin{center}
\includegraphics[angle=0,width=.6\linewidth,clip=true]{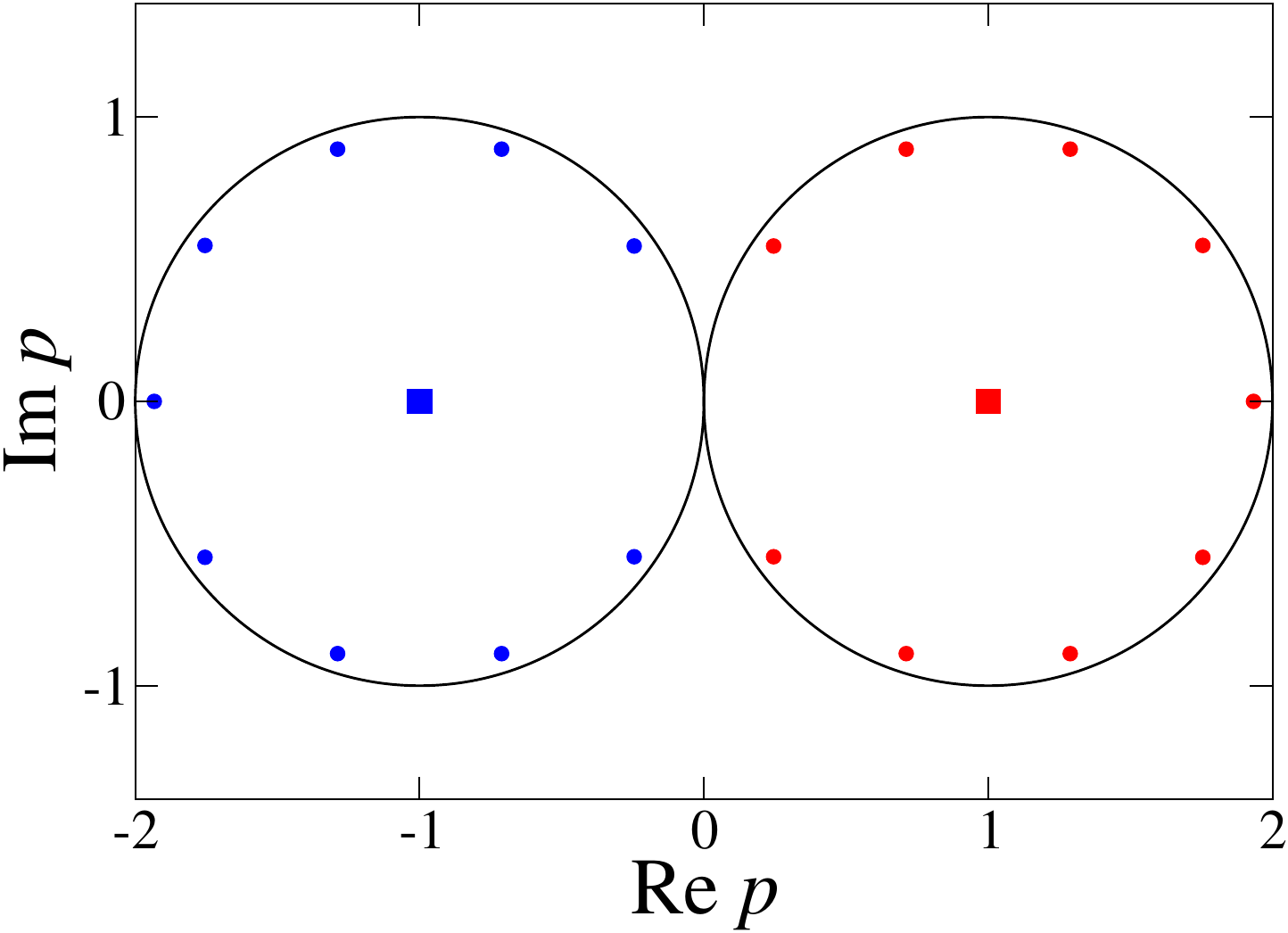}
\caption{
The 9 zeros $z_b$ (red symbols) and their opposites (blue symbols) in the complex $p$-plane for $M=10$.
The circles have unit radii and are centered at $p=\pm1$ (red and blue squares).
After~\cite{fpp}.}
\label{zerosplot}
\end{center}
\end{figure}

The family~\eqref{rhon} provides an interesting framework where a wide range of explicit results can be derived,
following the approach of~\cite{fpp}.
Another recent work~\cite{BMS} is devoted to the statistics of gaps
in random walks with the same family of step distributions.
These distributions belong to the class of distributions
whose Laplace transform $\hat\rho(p)$ is a rational function\footnote{As mentioned in~\cite{fpp},
the observation that Wiener-Hopf integral equations such as~(\ref{WH})
associated with such distributions are solvable by elementary means
dates back at least to the works of Wick~\cite{wick} and Chandrasekhar~\cite{chandra}.}.
The product formula\footnote{Here and throughout the following, sums and products over the zeros $z_b$
run over the range $b=1,\dots,M-1$, which is empty for $M=1$.}
\beq
\phi(p)=1-\hat\rho(p)=-\frac{p^2}{(1-p^2)^M}\prod_b(z_b^2-p^2)
\label{phiprod}
\eeq
plays a key role in the following.
In addition to a double zero at the origin,
$\phi(p)$ has $M-1$ zeros with positive real parts, denoted as $z_b$ for $b=1,\dots,M-1$, along
with their opposites.
These zeros sit near the circles with unit radii centered at $p=\pm1$ (see~Figure~\ref{zerosplot}).
They have been investigated in detail in~\cite{lfn}.
We have in particular
\beq
D=\frac{M(M+1)}{2}=\prod_bz_b^2.
\eeq

All relevant quantities can be expressed as symmetric functions of the zeros.
We have, e.g., (see~\eqref{eq:omegaint},~\eqref{eq:K},~\eqref{eq:ell})
\beq
\omega=M-\sum_bz_b,\qquad
K=2\prod_b(z_b-1),\qquad
\ell=M-\sum_b\frac{1}{z_b}.
\label{qties}
\eeq
The distributions of $H$ and $E$ read~\cite{fpp}
\beq
f_H(x)=\e^{-x}\sum_{k=0}^{M-1}(S_k-S_{k+1})\frac{x^k}{k!},\qquad
f_E(x)=\frac{\e^{-x}}{\sqrt{D}}\sum_{k=0}^{M-1}S_k\frac{x^k}{k!}.
\label{fhen}
\eeq
Using~(\ref{zhe}), we obtain the explicit expression
\beq
f_Z(x)=2\e^{-2x}\sum_{k,l=0}^{M-1}S_k(S_l-S_{l+1})\frac{x^{k+l}}{k!\,l!}.
\label{fzn}
\eeq
The density of the gap $Z$ is thus the product of the decaying exponential $2\e^{-2x}$
by a polynomial of degree $2(M-1)$.
In~(\ref{fhen}) and~(\ref{fzn}), the $S_k$ are the elementary symmetric functions of the variables
$z_b-1$,
defined by the identity
\beq
P_M(y)=\prod_b(1+(z_b-1)y)=\sum_{k=0}^{M-1}S_ky^k.
\label{sofy}
\eeq
We have thus
\beq
S_0=1,\qquad
S_1=\sum_b(z_b-1),\quad
\dots,\quad
S_{M-1}=\prod_b(z_b-1),\quad
S_M=0.
\eeq
The moments of the gap distribution are investigated in the Appendix.
Closed-form expressions for $\mean{Z^2}$ and $\mean{Z^3}$ are given in~(\ref{z2n}) and~(\ref{z3n}).

Explicit results can be given for the first few values of the integer $M$.

\begin{enumerate}

\item[1.]
For $M=1$,
the step distribution is the symmetric exponential distribution
(or Laplace distribution):
\beq
\rho(x)=\frac{\e^{-\abs{x}}}{2}.
\label{rho1}
\eeq
There are no zeros, so that we readily recover the simple
results~\cite{gmsprl,revue,mounaix1,mounaix2}
\beq
f_H(x)=\e^{-x},\qquad f_Z(x)=2\e^{-2x}.
\label{exps}
\eeq
We thus have
\beq
\omega=1,\quad K=2,\quad
\mean{H}=1,\quad \mean{H^2}=2,\quad \mean{Z}=\frac12,\quad \mean{Z^2}=\frac12.
\eeq

\item[2.]
For $M=2$,~(\ref{rhon}) reads
\beq
\rho(x)=\frac{\abs{x}\e^{-\abs{x}}}{2}.
\eeq
There is a single zero, $z_1=\sqrt{3}$.
Using~(\ref{fhen}) and~(\ref{fzn}), we recover~\cite{gmsprl,revue,mounaix1,mounaix2}
\beqa
f_H(x)
&=&\left(2-\sqrt{3}+(\sqrt{3}-1)x\right)\e^{-x},
\nonumber\\
f_Z(x)
&=&2\left(2-\sqrt{3}+(4\sqrt{3}-6)x+(4-2\sqrt{3})x^2\right)\e^{-2x}.
\eeqa
We thus have
\beqa
&&\omega=2-\sqrt3,\qquad K=2(\sqrt3-1),
\nonumber\\
&&\mean{H}=\sqrt3,\qquad \mean{H^2}=4\sqrt3-2,\qquad \mean{Z}=1,\qquad
\mean{Z^2}=\frac{5-\sqrt3}{2}.
\eeqa

\item[3.]
For $M=3$,~(\ref{rhon}) reads
\beq
\rho(x)=\frac{x^2\e^{-\abs{x}}}{4}.
\eeq
There is a pair of conjugate complex zeros, $z_1=a+\ii b$ and $z_2=a-\ii b$, with
\beq
a=\frac{\sqrt{2\sqrt6+3}}{2}\approx1.405256,\qquad
b=\frac{\sqrt{2\sqrt6-3}}{2}\approx0.689017,
\eeq
thus~(\ref{fhen}) and~(\ref{fzn}) yield
\beqa
f_H(x)
&=&\Bigl(3-2a+(4a-3-\sqrt6)x+(1+\sqrt6-2a)\frac{x^2}{2}\Bigr)\e^{-x},
\nonumber\\
f_Z(x)
&=&(a_0+a_1x+a_2x^2+a_3x^3+a_4x^4)\e^{-2x},
\eeqa
with
\beqa
a_0&=&6-4a,\qquad
a_1=28a-24-6\sqrt{6},\qquad
a_2=31+18\sqrt{6}-(38+6\sqrt{6})a,
\nonumber\\
a_3&=&(16+8\sqrt{6})a-20-12\sqrt{6},\qquad
a_4=5+2\sqrt{6}-(2+2\sqrt{6})a.
\eeqa
We thus have
\beqa
&&\omega=3-2a,\qquad K=2(1+\sqrt6-2a),\qquad
\\
&&\mean{H}=\sqrt6,\qquad \mean{H^2}=6\sqrt6-4a,\qquad \mean{Z}=\frac32,\qquad
\mean{Z^2}=\frac{51-(2+6\sqrt6)a}{8}.
\nonumber
\eeqa

\end{enumerate}

\begin{figure}[!htbp]
\begin{center}
\includegraphics[angle=0,width=.6\linewidth,clip=true]{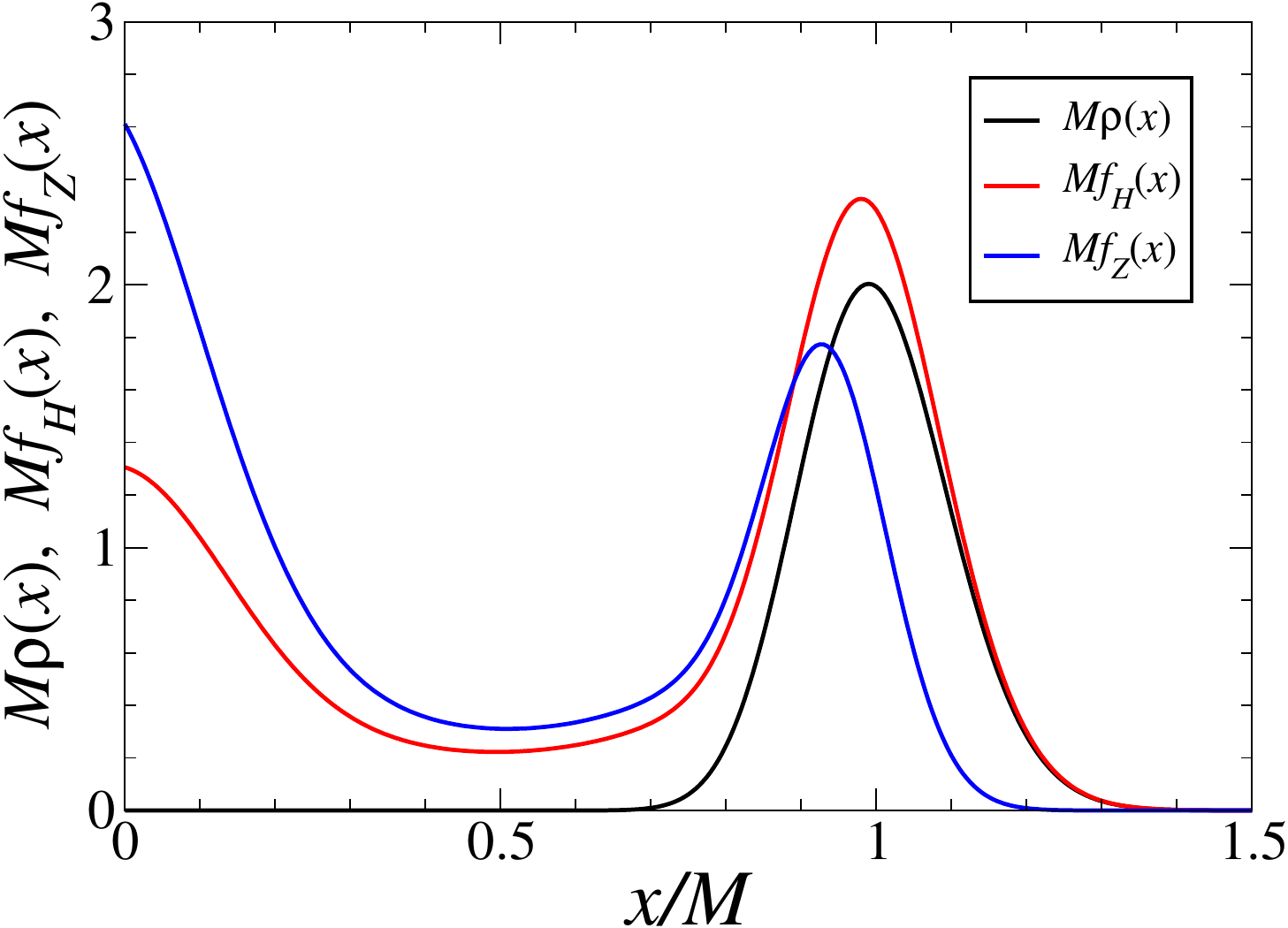}
\caption{
Comparison between
the positive part of the step distribution $\rho(x)$ (black),
the distribution $f_H(x)$ of the first positive position (red)
and the distribution $f_Z(x)$ of the topmost gap (blue) for $M=100$.
All distributions are rescaled by powers of $M$ for a better readability.}
\label{100plot}
\end{center}
\end{figure}

The distributions of $H$ and $Z$ exhibit remarkable scaling properties as the Erlang parameter $M$
gets large.
Figure~\ref{100plot} shows the positive part of $\rho(x)$,
and $f_H(x)$ and $f_Z(x)$ for $M=100$.
All distributions are rescaled by powers of $M$ for a better readability.
The step distribution $\rho(x)$ exhibits a narrow peak near $x/M=1$, as expected.
The distributions of $H$ and $Z$ exhibit a strongly bimodal character,
with a first peak for $x\ll M$, separated by a pronounced dip from a second narrower peak near
$x=M$,
resembling that of the step distribution $\rho(x)$.
This scenario was already described in~\cite{fpp} in the case of $H$.
It becomes more and more pronounced as the Erlang parameter $M$ is increased.

\begin{figure}[!htbp]
\begin{center}
\includegraphics[angle=0,width=.6\linewidth,clip=true]{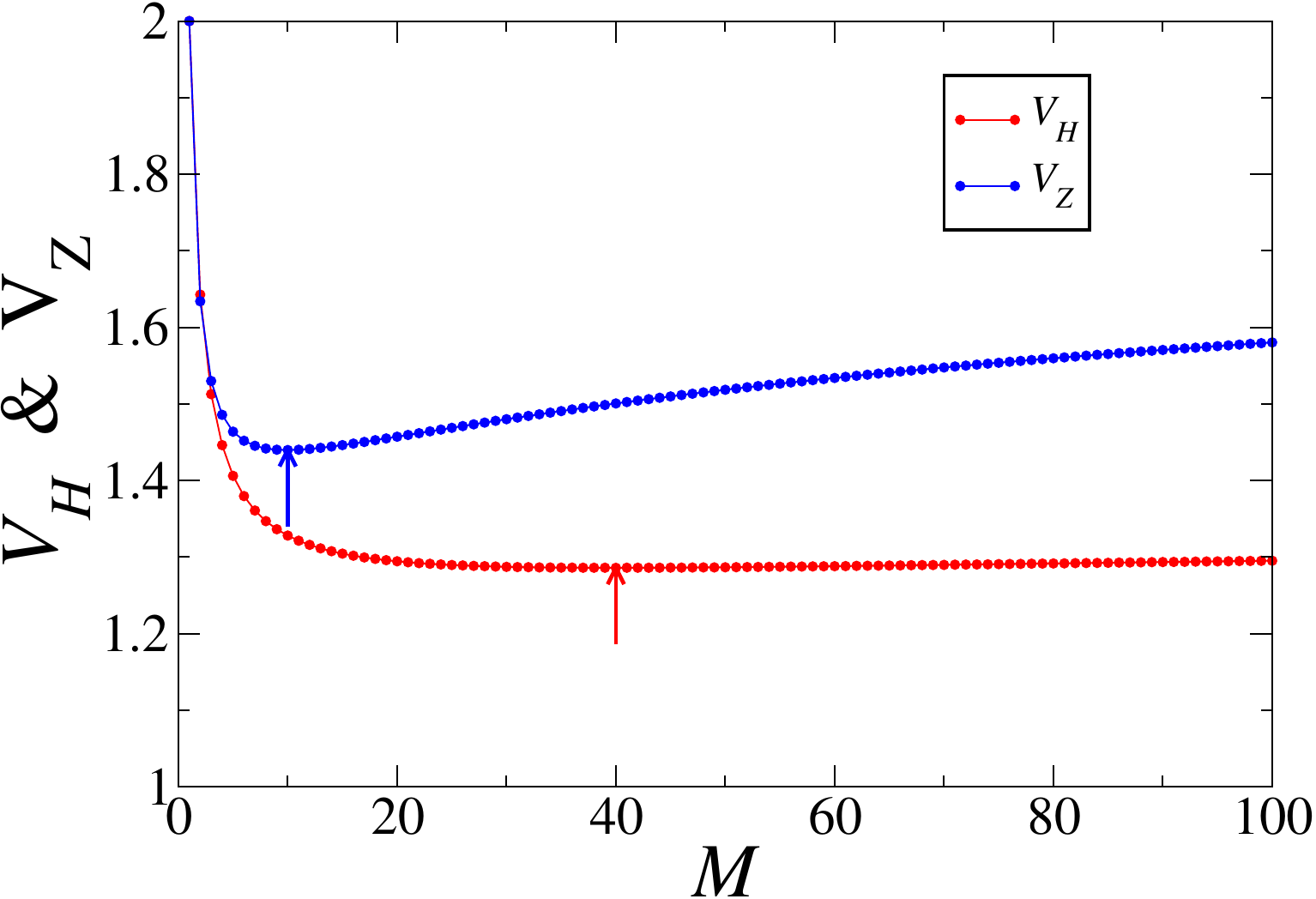}
\caption{
Moment ratios $V_H$ and $V_Z$, defined in~(\ref{vhvz}), against Erlang parameter $M$.
Arrows: minimal values of the moment ratios, respectively reached for $M=40$ and $M=10$.}
\label{vhvzplot}
\end{center}
\end{figure}

The above picture is corroborated and made quantitative by the large-$M$ estimates of moments
(see (\ref{momsn}), (\ref{h2sca}), (\ref{z2sca})):
\beqa
\mean{H}&\approx&\frac{M}{\sqrt2},\qquad
\mean{H^2}\approx\frac{M^2}{\sqrt2}\left(1-\frac{1}{\pi\sqrt{M}}\ln\frac{M}{M_0}\right),
\nonumber\\
\mean{Z}&=&\frac{M}{2},\qquad
\mean{Z^2}\approx\frac{M^2}{2}\left(1-\frac{2}{\pi\sqrt{M}}\ln\frac{M}{M_2}\right).
\label{hzmomsn}
\eeqa
The weights of the second peaks in the distributions of $H$ and $Z$ therefore respectively
equal $1/\sqrt2$ and $1/2$.
These numbers are consistent with~(\ref{evs}).
So, in the formal $M\to\infty$ limit, setting $y=x/M$, we have
\beqa
\rho(x)&\to&\frac12\left(\delta(y+1)+\delta(y-1)\right),
\nonumber\\
f_H(x)&\to&\left(1-\frac{1}{\sqrt2}\right)\delta(y)+\frac{1}{\sqrt2}\,\delta(y-1),
\nonumber\\
f_Z(x)&\to&\frac12\left(\delta(y)+\delta(y-1)\right).
\eeqa
Another way of looking at the problem is to consider the moments ratios (see~Figure~\ref{vhvzplot})
\beq
V_H=\frac{\mean{H^2}}{\mean{H}^2},\qquad
V_Z=\frac{\mean{Z^2}}{\mean{Z}^2}.
\label{vhvz}
\eeq
The first ratio was already considered in~\cite{fpp}, where it was denoted as~$2{\bf A}$.
Both ratios start at 2 for $M=1$, corresponding to the exponential distributions~(\ref{exps}),
decrease rather quickly, reach their minimal values $V_H\approx1.285921$ and $V_Z\approx1.439742$
at $M=40$ and $M=10$
respectively, and then converge very slowly to their limits, as obtained from~(\ref{hzmomsn}):
\beq
V_H\approx\sqrt2\left(1-\frac{1}{\pi\sqrt{M}}\ln\frac{M}{M_0}\right),\qquad
V_Z\approx2\left(1-\frac{2}{\pi\sqrt{M}}\ln\frac{M}{M_2}\right).
\eeq
The numerical values of $M_0$ and $M_2$ are given in~(\ref{nums}).

\appendix
\section{Gap moments for Erlang distributions}

By definition, these moments read
\beq
\mean{Z^n}=\int_0^\infty\dd x\,x^n\,f_Z(x).
\eeq
For the family of symmetric Erlang distributions~(\ref{rhon}), $f_Z(x)$ is given by~(\ref{fzn}).
Integrating term by term, we obtain
\beq
\mean{Z^n}=\sum_{k,l=0}^{M-1}S_k(S_l-S_{l+1})\frac{(k+l+n)!}{2^{k+l+n}\,k!\,l!},
\eeq
which can be rewritten as
\beq
\mean{Z^n}=n\sum_{k,l=0}^{M-1}S_kS_l\frac{(k+l+n-1)!}{2^{k+l+n}\,k!\,l!}.
\label{zn}
\eeq

\subsection{First moment}

To begin with, consider the first moment.
Equation~(\ref{zn}) yields
\beq
\mean{Z}=\sum_{k,l=0}^{M-1}S_kS_l\frac{(k+l)!}{2^{k+l+1}\,k!\,l!}.
\label{z11}
\eeq
Using the integral representation
\be
\frac{(k+l)!}{k!\,l!}=\int\frac{\dd u}{2\pi\ii}\,\frac{1}{u^{k+1}(1-u)^{l+1}}
\ee
along a vertical contour with $0<\Re u<1$,
the sums over $k$ and $l$ can be performed separately by using~(\ref{sofy}),
yielding
\be
\mean{Z}=\frac{1}{2}\int\frac{\dd u}{2\pi\ii\,u(1-u)}\,
P_M\left(\frac{1}{2u}\right)P_M\left(\frac{1}{2(1-u)}\right).
\ee
Setting $u=(1+p)/2$ and using the product formula~(\ref{phiprod}), we obtain
\be
P_M\left(\frac{1}{2u}\right)P_M\left(\frac{1}{2(1-u)}\right)=
\prod_b\frac{z_b^2-p^2}{1-p^2}=-\frac{1-p^2}{p^2}\,\phi(p),
\ee
which leads to
\beq
\mean{Z}=-\int\frac{\dd p}{2\pi\ii}\,\frac{\phi(p)}{p^2}.
\label{z12}
\eeq
This formula is identical to~(\ref{z1int}), which holds for arbitrary step distributions.
In the present case, using~(\ref{laprhon}), the contour integral in~(\ref{z12}) evaluates to
\be
\mean{Z}=\frac{M}{2},
\ee
in agreement with~(\ref{z1res}),~(\ref{momsn}).

\subsection{Second moment}

For the second moment,~(\ref{zn}) yields
\beq
\mean{Z^2}=\sum_{k,l=0}^{M-1}(k+l+1)S_kS_l\frac{(k+l)!}{2^{k+l+1}\,k!\,l!},
\label{z21}
\eeq
which differs from~(\ref{z11}) by the factor $(k+l+1)$.
Proceeding as above, the sums over $k$ and $l$ can be performed by using~(\ref{sofy}),
as well as
\beq
\sum_{k=0}^{M-1}kS_ky^k=y\frac{\dd}{\dd y}P_M(y)=\Sigma_1(y)P_M(y),
\label{sder1}
\eeq
with
\beq
\Sigma_1(y)=\sum_b\frac{(z_b-1)y}{1+(z_b-1)y}.
\eeq
We finally obtain
\beq
\mean{Z^2}=-\int\frac{\dd p}{2\pi\ii}\,
\frac{\phi(p)}{p^2}\left(1+2\sum_b\frac{z_b(z_b-1)}{z_b^2-p^2}\right).
\label{z22}
\eeq
This contour integral evaluates to
\beq
\mean{Z^2}=M\left(\ell-\frac12\right)
-\sum_b\frac{z_b-1}{z_b^2}\left(1-(1+z_b)^{-M}\right).
\label{z2n}
\eeq

\subsection{Third moment}

For the third moment,~(\ref{zn}) yields
\beq
\mean{Z^3}=\frac{3}{4}\sum_{k,l=0}^{M-1}(k+l+1)(k+l+2)S_kS_l\frac{(k+l)!}{2^{k+l+1}\,k!\,l!}.
\label{z31}
\eeq
Proceeding as above, the sums over $k$ and $l$ can be performed by
using~(\ref{sofy}),~(\ref{sder1}), and
\beq
\sum_{k=0}^{M-1}k^2S_ky^k=\left(y\frac{\dd}{\dd y}\right)^2P_M(y)
=\left(\Sigma_1(y)^2+\Sigma_2(y)\right)P_M(y),
\eeq
with
\beq
\Sigma_2(y)=y\frac{\dd}{\dd y}\Sigma_1(y)=\sum_b\frac{(z_b-1)y}{(1+(z_b-1)y)^2}.
\eeq
Skipping details, we are thus left with
\beq
\mean{Z^3}=-\frac{3}{2}\int\frac{\dd p}{2\pi\ii}\,\frac{\phi(p)}{p^2}
\left(1+\sum_b\frac{(z_b-1)(5z_b-1)}{z_b^2-p^2}
+2\sum_{b\ne c}\frac{z_bz_c(z_b-1)(z_c-1)}{(z_b^2-p^2)(z_c^2-p^2)}\right).
\label{z32}
\eeq
This contour integral evaluates to
\beqa
\mean{Z^3}
&=&\frac{M^2(M+5)}{16}+\frac{3M}{2}\left(\ell-\frac12\right)^2
\nonumber\\
&-&\frac{3}{4}\sum_b\frac{(z_b-1)(5z_b-1)}{z_b^3}\left(1-(1+z_b)^{-M}\right)
\label{z3n}
\\
&+&\frac{3}{2}\sum_{b\ne c}\frac{(z_b-1)(z_c-1)}{z_b^2-z_c^2}
\left(\frac{z_c}{z_b^2}\left(1-(1+z_b)^{-M}\right)
-\frac{z_b}{z_c^2}\left(1-(1+z_c)^{-M}\right)\right).
\nonumber
\eeqa

The general structure of the gap moments appears clearly from~(\ref{z2n}) and~(\ref{z3n}).
The expression of $\mean{Z^n}$ involves,
in addition to an explicit polynomial in the Erlang parameter $M$
and the extrapolation length $\ell$,
a simple sum over the zeros $z_b$,
a double sum, and so on, up to a sum over $(n-1)$-uples of different zeros.

\subsection{Large-$M$ asymptotics}

The asymptotic behaviour of symmetric functions of the zeros $z_b$
can be analysed using an approach introduced in~\cite{lfn}.
As the parameter $M$ increases,
the zeros become uniformly distributed along a circle centered at $p=1$
with a radius close to unity, given by
\beq
z_b\approx1-2^{-1/M}\e^{-2\pi\ii b/M}\qquad(b=1,\dots,M-1).
\label{zuni}
\eeq
This leading-order estimate is however not accurate enough in most practical cases.
For a large but finite $M$, the positions of the zeros near the origin, i.e., for $b\ll M$ or
$M-b\ll M$,
deviate significantly from the uniform distribution mentioned above.
To be specific, let us consider the range $b\ll M$.
Retaining the notations of~\cite[Sec.~6.2]{lfn}, and defining
\be
\xi=\frac{2\pi b}{\sqrt{M}},
\ee
the zeros $z_b$ satisfy the scaling formula
\beq
z_b\approx\frac{2\pi\ii b+Y(\xi)}{M},
\label{zsca}
\eeq
where $\cosh Y(\xi)=\exp(\xi^2/2)$, i.e.,
\be
Y(\xi)=\frac{\xi^2}{2}+\ln\left(1+\sqrt{1-\e^{-\xi^2}}\right).
\ee

For the extrapolation length $\ell$ (see~(\ref{qties})),
some algebra yields~\cite{lfn,fpp}
\beq
\ell\approx\frac{M}{2}\left(1-\frac{1}{\pi\sqrt{M}}\ln\frac{M}{M_0}\right),
\label{ellsca}
\eeq
and therefore (see~(\ref{h12res}))
\beq
\mean{H^2}\approx\frac{M^2}{\sqrt2}\left(1-\frac{1}{\pi\sqrt{M}}\ln\frac{M}{M_0}\right),
\label{h2sca}
\eeq
with
\beq
M_0=4\pi^2\e^{-2I_0},\qquad
I_0=\int_0^\infty\frac{\dd\xi}{\xi^2}\,\left(\ln\left(1+\sqrt{1-\e^{-\xi^2}}\right)-\xi\e^{-\xi}\right).
\eeq
Applying the same to the sum entering~(\ref{z2n}) gives
\beq
\sum_b\frac{z_b-1}{z_b^2}\left(1-(1+z_b)^{-M}\right)
\approx\frac{M^{3/2}}{2\pi}\,\ln\frac{M}{M_1},
\label{sig1sca}
\eeq
with
\be
M_1=4\pi^2\e^{-2I_1},\qquad
I_1=\int_0^\infty\frac{\dd\xi}{\xi^2}\,\left(\sqrt{1-\e^{-\xi^2}}-\xi\e^{-\xi}\right).
\ee
By inserting the estimates~(\ref{ellsca}) and~(\ref{sig1sca}) into~(\ref{z2n}), we obtain
\beq
\mean{Z^2}\approx\frac{M^2}{2}\left(1-\frac{2}{\pi\sqrt{M}}\ln\frac{M}{M_2}\right),\qquad
M_2=4\pi^2\e^{-I_0-I_1}.
\label{z2sca}
\eeq
A numerical evaluation of the above constants yields
\beqa
&&I_0\approx0.787841,\quad
I_1\approx1.416383,
\nonumber\\
&&M_0\approx8.166752,\quad
M_1\approx2.323296,\quad
M_2\approx4.355890.
\label{nums}
\eeqa

\addcontentsline{toc}{section}{References}

\bibliography{papergap.bib}

\end{document}